\documentstyle[11pt,aasms4]{article}

\slugcomment{Submitted to Astrophysical Journal Lett.}
\begin{document}
\baselineskip 0.33in 
\title{STAR CLUSTERS AND THE STRUCTURE OF THE ISM.
TUNNELS AND WAKES IN GIANT EXTRAGALACTIC H II REGIONS}

\author{Guillermo
Tenorio-Tagle}
\affil{Instituto Nacional de Astrof\'\i sica Optica y Electr\'onica,
AP 51, 72000 Puebla, M\'exico}

\author{Gustavo A. Medina-Tanco$^{1,2}$}

\affil{ 1.
Instituto Astron\^omico e Geof\'{\i}sico, University of S\~ao Paulo, Brasil
\\
gustavo@iagusp.usp.br
}

\affil{ 2. Dept. of Physics and Astronomy, University of Leeds,
Leeds LS2 9JT, UK}

\abstract{
Several structures have been discovered embedded in regions of recent
or ongoing star formation, which point to the importance of the
interaction between fast moving wind-blowing stars and their environment.
Using hydrodynamic simulations, we investigate the passage through the
interstellar medium of a supersonic stellar wind source, and show how
it can naturally lead to the formation of tubes, channels and swamps
of globules as interfaces are crossed. The results are in excellent
agreement with observation of 30 Doradus.
}

\subjectheadings{{\bf ISM}: kinematics and dynamics -- H {\sc ii} regions }

\section{INTRODUCTION}

Prompted by the rich structure detected in
 extragalactic H {\sc ii } regions, as that recently captured by HST
in 30 Doradus (see Hunter et al. 1995a, and
Scowen et al. 1998), and the strong resemblance that
these features seem to
have with those found in Galactic
H {\sc ii} regions, we have started a series of
numerical hydrodynamical calculations aimed at unveiling the physical
parameters
that define the structure of interstellar matter in
regions of massive star formation.  Among the most noticeable features are
the
pillars, fingers, and globules, all apparently immersed in  the
hot coronal phase of the ISM,  as well as the channels or
tubes most likely blown by
the wind of rapidly moving stars. Most of these features
appear to be pointing towards the central
stellar cluster of 30 Doradus;
similarly to what happens in Galactic HII regions such as in M16,
Orion, etc (Bok et al. 1970).

The 30 Doradus nebula is powered by NGC 2070,
a 2 - 4 $\times 10^6$ yr old massive ($M_*$ $\sim$ a few  $\times 10^5$ $M_{\odot}$)
starburst
(Hunter et al. 1995b) with more than 300 OB stars, and
an incipient population of
Wolf - Rayet stars, all of them producing an intense UV flux
and energetic stellar winds. The energetics from the cluster are
most likely responsible for the clearing of matter around
the central region which presents very little nebulosity and is
surrounded by a
bright circular rim of about 5 pc in radius.
This is similar to other structures: extended bright arcs, rims or ridges,
seen at larger radii within 30 Doradus (see Melnick et al. 1998).
All of these present strong density fluctuations and
seem to be crowded with large numbers of pillars fingers
and globules sticking out and
pointing towards the stars. Hunter et al. (1995a) and
Scowen et al. (1998) have also displayed other interesting
structures in 30 Doradus and among these, perhaps the most striking is the
half a pc long parabola-shaped structure in the ridge of emission to
the west of the central cluster. The parabola opens away from the cluster
and from
the AOI star located 2 pc from the centre. Also, at the focus of the
parabola,
some 0.13 pc from the arc, is a small non-stellar knot, perhaps associated
with
the protostar candidate (P3) of Hyland et al. (1992). Similar to this is
the   1.5 pc long tube-like structure,
E - W oriented in the WF2 field of Scowen et al. (1998, their 
figures 2 and 7)
HST observation.
The structure is some 15 pc away from the central cluster
and seems embedded into a low emission background cloud.
It appears most evident in [SII] and it is also well detected
in H$\alpha$, but hardly in
[OIII]. This latter fact could be an indication of
a certain degree of inclination
into the cloud and  thus, the observed size may be a lower limit to
the true size of the (projected) structure.
The end of the structure, the one  closest to the cluster, is
very fragmented  and more circular in appearance with a radius of 7 $\times
10^{17}$ cm, while along its length the cylindrical structure presents a
radius $\sim$ 3 $\times 10^{17}$ cm. The inner part of the
tube must have a low density gas and contributes little to the
line emission. However, its walls present  a relatively
large density contrast which makes them evident relative to the
background nebula. Limb brightening along the walls may exaggerate their
actual thickness which was measured by
Scowen et al. to be about $10^{17}$ cm.
Also striking is the presence of a star at the centre of the fragmented end
of the tube, which led to the suggestion that
a high proper motion star that ploughs
through  a cloud of intermediate density while undergoing a stellar wind,
could be the agent causing the unusual structure (see Scowen et al. 1998).
Here we follow this suggestion and
investigate
the structure produced in the ISM
by the supersonic passage of a stellar wind source.

Section 2 is devoted to a two dimensional numerical study of the bow shocks
and wakes generated in the ISM
by the supersonic passage of wind-blowing stars while propagating
into either a constant density medium and into a steep density
discontinuity. Section 3 summarizes our conclusions.

\section{Stellar winds, bow shocks and wakes}

The interaction of a strong stellar wind with the
surrounding interstellar medium, is well known to
lead to an outer shock wave that accelerates and collects the
ISM into a dense shell,  while a reverse, or inner, shock thermalizes the
stellar wind gas (Weaver et al. 1977). Interior to the inner shock
one finds the free-wind region, characterized by a density distribution that
decreases outwards as $R^{-2}$, and  a
maximum, or terminal,  velocity of the ejected wind.
Upon deceleration at the inner shock, the
stellar wind thermal pressure becomes the piston that drives the outer shock
to incorporate more matter in the dense outer shell. The stellar
wind configuration is strongly modified if the star moves
with a velocity larger than the sound speed of the
thermalized wind ($v_{*} > c_{\mbox{\it shocked wind\/}}$).
In such a case, the flow
pattern is rapidly modified as the outer shock
approaches a bow shock configuration, supported by the wind ram
pressure (see Weaver et al.).  Note however, that winds from massive stars
($v_{wind}$ $\geq$ 1000 km/sec), as invoked by Scowen et al., lead to
thermalized winds with a high sound speed ($c_{\mbox{\it shocked wind\/}}
\geq$ 350 km/sec) and thus
for such stars to cause a bow shock, they must move  with even larger
speeds; larger than their computed value of 230 km/sec.

Rapidly moving stars in regions of recent star formation naturally result
from all of those
in eccentric orbits around the centre of mass
of the newly formed stellar cluster. Bound stars that traverse the region
all the way up to the cluster tidal radius ($\sim$ 100 pc) must reach
velocities
along their tracks of up to several tens of km/sec. It therefore seems
that less energetic winds; either a slow wind from a massive star or
one similar to those expected from low-mass
stars ($v_{wind} \leq 100$ km/sec), are better candidates to produce
bow shocks in the HII region, as they plough through the clouds
interspersed
along their orbits.


\subsection{The Calculations}

All simulations were carried out with an explicit Eulerian two-dimensional
    hydrodynamic code (see R\'o\.zyczka 1985). The code is written in
    cylindrical coordinates with symmetry about the $z$ axis and at the
    equator and thus calculates only one quadrant of the flow.  The
    code was adapted to account for a steady wind for which
    one could prescribe both the mass loss rate
    ($\dot{M}$) and its terminal velocity
($v_{wind}$) as well as for the motion of the star ($v_*$).
      The calculations were performed in a grid of
    900 by 250 cells, evenly spaced and with a resolution of
$\sim 10^{-3}$ pc.
In all cases radiative
cooling was taken into consideration by means of the integration
of the standard cooling function (Raymond, Cox \& Smith 1976).

Figure 1 compares two calculations which only differ in the assumed
temperature value of the background gas. For case a (Figure 1a) a
cold medium of
temperature T$_0$
= 100 K was assumed, while for case b (Figure 1b) we assumed
a photoionized environment
and thus a T$_0$ = 10$^4$ K. Otherwise, the stellar mass-loss rate
$\dot{M}$ (=10$^{-7}$ $M_{\odot}$ yr$^{-1}$) and a wind terminal speed , $v_{wind}$
(=50 km/sec) are the same in both cases.
Also, the density of the background gas ($\rho_0$ = 10$^{-23}$ gr
cm$^{-3}$) and the
velocity of the star ($v_*$ = 30 km/sec) are identical in both calculations.
In both cases $v_*$ is supersonic with respect to the shocked wind and
to the background gas,
 causing the formation of a bow shock
into the ISM  with Mach numbers $\sim$ 30 and 3, respectively.
Clearly in case a, the lower temperature gas allows the bow shock to span,
as a supersonic disturbance, across a much larger volume than in case b.
Note that the affected radius ($r(z)$) grows with the distance, $z$,
behind the star in excellent agreement with
 the analytic solution of Weaver et al. (1977):

\begin{equation}
r(z) = (10 M \dot{v}_{wind}^2 / (33 \pi \rho_0 v_*^3))^{0.25}z^{0.5}
\end{equation}

In case b however,
the oblique section of the shock soon become subsonic and
thus the
passage of the rapidly moving wind-blowing star affects a smaller volume of
the background cloud. In both numerical calculations however, the
stellar wind ram
pressure

\begin{equation}
\rho_{wind}v_{wind}^2 = \dot{M}_{wind} v_{wind}/(4 \pi R_{shock}^2)
\end{equation}

\noindent balances the ram pressure of the background gas

\begin{equation}
P_{ram} = \rho_0 v_*^2
\end{equation}

\noindent at a distance

\begin{equation}
R_{shock} = (\dot {M}_{wind} v_{wind}/(4 \pi \rho_0  v_*^2)
)^{0.5}
\end{equation}

\noindent and it is there, at $R_{shock}$, where the wind meets
a reverse shock and becomes thermalized. Note that
if $v_*$ is larger than the sound speed of the thermalized wind, 
then a bow shock will 
develop (Weaver et al. 1977). In such a case, multiple 
combinations  of $\dot{M}$, $v_{wind}$, $\rho_0$ and $v_*$
may lead to the same values of $R_{shock}$ and to similar
bow shock patterns and thus to a similar expected impact on the background gas.
The calculations shown in Figure 1 thus imply that it is the
ram pressure of the background gas and not its thermal
pressure, as assumed by Scowen et al. (1998),
what defines the flow pattern. In both solutions once the wind
becomes thermalized it streams towards the back of the flow and
expands to evenly fill the
cavity (or inner volume of the tube) left by the passage of the bow shock,
which presents a radius similar to $R_{shock}$.
There is a well established density contrast, of more than
an order of magnitude, between the shocked wind matter filling the
inner volume of the tube and the walls of displaced and
accelerated shocked ISM, and the density contrast is even
larger at the head of the configuration. We also note
that both flow patterns are stable against fragmentation
(see Figures 1a, b).

The same stellar wind parameters, stellar velocity,
background density and temperature as in case b, have been
used in a calculation aimed at
establishing the physical properties of the interstellar tube feature
recently found in the Tarantula nebula (Scowen et al. 1998).
In this case however, a large density discontinuity was placed some 1 pc
ahead of the stellar
wind source. The sharp discontinuity leads to values of density
($n_{coronal}$ = 0.1 cm$^{-3}$) and temperatures ($T_{coronal}$ = 10$^6$ K),
representative of a coronal phase in pressure equilibrium with the
externally photoionized cloud into which the star originally moves.
The coronal phase parameters so established are in excellent agreement
with the values derived from Wang \& Helfand (1991)
from their X-ray observations of
30 Doradus.

Figure 2 shows the flow patterns produced as the
star approaches and enters the coronal medium. The time sequence shows
how the bow shock evolves into a subsonic disturbance and stops sweeping gas
as soon as it comes into contact with the hot coronal phase.
At the same time, the
inner shock moves (in agreement with relation 4)
to a much larger radius, allowing the free-wind
to occupy a larger volume.

The layer of shocked gas around the leading bow shock then
rapidly begins to fragment into a collection
of dense condensations, or globules, driven away from the star,
by conservation of their own
momentum, causing the appearance of
a more spherical end of the elongated remnant.
The last panels of Figure 2 shows a long cylindrical channel
(almost one pc long), with a fragmented end that presents a cross-section
 almost twice as large as the full thickness of the sector of the
tube drilled within the cloud,  in  good agreement with the observations.

Figure 3 shows a more advanced stage in the evolution when the fragmented
globules have managed to fully detach from each other, while  moving
subsonically into the pervasive coronal phase. The globules are also
being left behind the star which remains
undecelerated,
while its wind is now confined by the thermal pressure of the coronal
medium.

\section{CONCLUSIONS}

Our numerical simulations show that parabola-shaped and
tube-like structures such as those found in 30 Doradus,
could indeed be generated by the rapid passage of
wind-blowing stars.
These objects  are to become supersonic as they follow their bound trajectories
around the newly formed cluster.
We have shown that their possible
broken end appearance could result when entering a large density
discontinuity.
The latter seems fully justified by
the fact that clouds appear to be
immersed into a hot coronal medium, and thus upon crossing the
discontinuity, fragmentation of the bow shocked matter occurs.

Several conclusions can be drawn from our calculations.

1) Tubes or channels in the ISM would result
from the rapid passage of stars undergoing  winds, and thus presenting
enhanced cross-sections, if they plough supersonically
through their shocked wind and the ionized ISM.

2) A supersonic passage sets a bow shock that displaces and condenses the
overtaken material, while the thermalized wind evenly fills, with a low
density gas, the inner sector of the leftover channel structure.

3) The typical dimension (width) of the channels
generated by such
events,  is determined  by
the balance between the stellar wind ram pressure and that exerted
by the background gas (see eq 4).

Regarding the tube structure seen in 30 Doradus (Scowen et al. 1998), we have
shown that:

4) Such structures can arise from many combinations of the
stellar wind parameters,
background densities and stellar velocities (see eq 4),
including among the latter
supersonically moving stars traveling in eccentric bound orbits around
NGC 2070, causing the resultant structure to point
towards the central cluster.

5) Our calculations also indicate that the wider and
fragmented end presented by the
tube structure found in 30 Doradus, may naturally develop as  the
bow shock  pattern crosses the discontinuity between the
dense cloud and the pervasive coronal phase, that seems to permeate
all clouds in the neighborhood of the centre of 30 Doradus.

6) The fragmentation of the bow shock pattern was also shown to lead,
during the late stages of the calculations, to a swamp of globules immersed
in an extended low density free-wind region.

\vskip 3.0cm

Our thanks to Roberto and Elena
Terlevich, Sally Oey and Casiana Mu\~noz-Tu\~non for
many interesting discussions. This work was partially supported by
the Brazilian agency FAPESP.

\vfil\eject

{\bf Figure Captions}
\figcaption{ Bow shocks and wakes.
The panels shows isodensity contours and the velocity field (indicated with
the length of the arrows proportional to speed) produced by the
supersonic motion of the stellar wind source.
The upper and lower panels are snapshots at $t$ = $5.3 \times 10^{4}$ yr,
and
$2.6 \times 10^4$ yr, respectively. In both cases the flow has reach its
steady state configuration.
  The contours are
logarithmically spaced with $\Delta$ log$ \rho =0.2$.
The distance between consecutive tick marks is $6.5 \times 10^{17}$ cm
in the upper panel and $4.9 \times 10^{17}$ cm in the lower one as they
were calculated in different grids. Note however, that the distance
$R_{shock}$ is the same in both cases.
In both cases the
maximum central wind speed is 50 km/sec.}

\figcaption{Tubes and blowout in the ISM.
Time sequence showing the
isodensity contours and the velocity field (indicated with
the length of the arrows proportional to speed) produced by the
supersonic motion of the stellar wind source.
The panels are snapshots at $t$ = $1.8, 2.6, 3.4, 4.1, 4.8, and 5.5
\times 10^{4}$ yr.
  The contours are
logarithmically spaced with $\Delta$ log$ \rho =0.2$.  The distance
between consecutive tick marks is $4.9 \times 10^{17}$ cm.  The
maximum central wind speed is 50 km/sec. }

\figcaption{An isolated stellar wind source immersed in the pervasive
coronal gas.
The same as Figure 2 for the last calculated time
$t$ = $6.0
\times 10^{4}$ yr.}

\vfil\eject
\references

\noindent Bok, B. J., Cordwell, C. S. \& Cromwell, R.: 1970, in Symposium on
Dark Nebulae,
Globules and Protostars. University of Arizona Press

\noindent Hunter, D. A., Shaya, E, J., Scowen, P., Hester, J. J.,
Groth, J. G., Lynds, R. \&
O'Neil, E. J.: 1995, ApJ 444, 758

\noindent Hunter, D. A., Shaya, E, J., Holtzman, J. A., Light, R. M.,
O'Neil, E. J., \& Lynds, R.: 1995, ApJ 448, 179

\noindent Hyland, A. R., Straw, R., Jones, T. J. \& Gatley, I.: 1992, MNRAS
257, 391

\noindent Melnick, J., Tenorio-Tagle, G. \& Terlevich, R.: 1998, MNRAS
(in press)

\noindent R\'o\.zyczka, M.: 1985, A\&A, 143, 59

\noindent Raymond, J. C., Cox, D. P. \& Smith, B. W.:  1976, ApJ,  204, 290

\noindent Scowen, P. A., Hester, J. J., Sankrit, R., Gallagher, J. S.,
Ballester, G. E., Burrows, C. J., Clarke, J. T., Crisp, D., Evans, R. W.,
Griffiths, R. E., Hoessel, J. G., Holtzman, J., Krist, J. R., Mould, J. R.,
Stapelfeld, K. R.,
Trauger, J. T., Watson, A. M., \& Westphal, J. A.: 1998, AJ (in press)


\noindent Wang, Q. \& Helfand, D. J.: 1991, ApJ, 370, 541

\noindent Weaver, R., McCray, R., Castor, J., Shapiro, P. \& Moore, R.: 1977
ApJ, 218, 377

\endreferences

\vfil\eject

\end{document}